\documentclass[a4paper,11pt]{article}

\usepackage{jinstpub} 

\graphicspath :
\graphicspath{{Figures/}}

\usepackage{xcolor}

\title{\boldmath Pruning: a tool to optimize the layout of large scale arrays for ultra-high-energy air-shower detection}

\author[a,c]{Aur\'elien Benoit-L\'evy,}
\author[b,d,c]{Kumiko Kotera,}
\author[e,f]{Matías Tueros}

\affiliation[a]{Universit\'{e} Paris-Saclay, CEA, List, F-91120, Palaiseau, France}
\affiliation[b]{Sorbonne Universit\'{e} et CNRS, UMR 7095, Institut d'Astrophysique de Paris, 98 bis bd Arago, 75014 Paris, France}
\affiliation[c]{Department of Physics, Pennsylvania
State University, University Park, PA, USA}
\affiliation[d]{Vrije Universiteit Brussel, Physics Department, Pleinlaan 2, 1050 Brussels, Belgium}
\affiliation[e]{IFLP - CCT La Plata - CONICET, Casilla de Correo 727 (1900) La Plata, Argentina}
\affiliation[f]{Depto. de F\'isica, Fac. de Cs. Ex., Universidad Nacional de La Plata, Casilla de Coreo 67 (1900) La Plata, Argentina}

\emailAdd{aurelien.benoit-levy@cea.fr}
\emailAdd{kotera@iap.fr}
\emailAdd{tueros@fisica.unlp.edu.ar}

\abstract{The deployment of several large scale arrays is envisioned to study astroparticles at ultra-high energies. In order to circumvent the heavy computational costs of exploring and optimizing their layouts, we have developed a pruning method. It consists in i) running a set of microscopic simulations and interpolate them over a dense, regularly spaced array of detection units, and ii) pruning the unnecessary units out of the layout, in order to obtain the shower footprint on a newly shaped layout. This method offers flexibility to test various layout parameters, instrumental constraints, and physical inputs, with a drastic reduction in the required CPU time. The method can be universally applied to optimize arrays of any size, and using any detection techniques. 

For demonstration, we apply the pruning tool to radio antenna layouts, which allows us to discuss the interplay between the energy and inclination of air-showers on the size of the radio footprint and the intensity of the signal on the ground. Some rule-of-thumb conclusions that can be drawn for this specific case are: i) a hexagonal geometry is more efficient than a triangular geometry, ii) the detection efficiency of the array is stable to changes in the spacing between radio antennas around 1000$\rm{m}$ step size, iii) for a given number of antennas, adding a granular infill on top of a coarse hexagonal array is more efficient than instrumenting the full array with a less dense spacing. 
}

\begin{document}
\maketitle
\flushbottom

\section{Introduction}

Because of their intrinsic low fluxes, the observation of ultra-high-energy (UHE) cosmic particles requires the deployment of efficient detection units over gigantic areas. The next-generation ground experiments such as GRAND~\cite{GRAND20} or GCOS~\cite{GCOS_ICRC} are envisioning to deploy tens to hundreds of thousands of detection units over tens to hundreds of thousands of square kilometers. 
The geometrical layout of these units has to be designed to optimize the triggering rate, folding in the characteristics of the reconstruction methods and the demands of the science case. 

The exploration of layout geometries requires to run massive air-shower simulations scanning various configurations and corresponding parameter spaces. For regular layouts with  repeating patterns, the impact of the characteristics of each single pattern (geometry, size, etc.) has to be assessed. More complicated patterns with different granularity (logarithmic, fractal, with denser infills, etc.) can also be explored for specific scientific purposes. 

In order to reduce this demanding computational cost, which is usually the bottleneck to perform these type of studies, we have developed a {\it pruning} method. It consists in running a set of microscopic simulations and interpole them over a dense, regularly spaced array of detection units, and prune the unnecessary units out of the layout, in order to obtain the radio footprint on a newly shaped layout. By construction, the spacing between the detection units in the new layout has to be proportional to the step size of the initial dense array.

The pruning method can be universally applied to optimize arrays of any size, and using any detection technique (scintillators, \v{C}erenkov water tanks, radio antennas, ...). In this paper, for demonstrative purposes, we focus on the radio-detection of UHE particles arriving on Earth with inclined zenith angles. We illustrate the method on mid-scale (typically 100 radio antennas over 100 km$^2$) and small scale (a dozen antennas over less than 10 km$^2$) layouts, where applying the pruning tool will allow us to discuss the effects of the different layout geometries and of the instrumental (triggering) constraints on the detectability of UHE cosmic rays. 

In Section~\ref{section:methodology}, we outline the methodology adopted to implement the pruning tool and to test it over a set of ultra-high energy cosmic ray simulations. In Section~\ref{section:exploring}, we apply the pruning method on a radio antenna array, and present how it is possible to assess the impact of various key parameters characterizing a layout. We draw our conclusions and perspectives for using this tool in Section~\ref{section:conclusion}.

We illustrate these effects further with the computation of effective areas and event rates over selected layout geometries using the TALE cosmic-ray spectrum \cite{2018ApJ...865...74A}.

\section{Methodology}\label{section:methodology}

While entering the atmosphere, UHE particles produce extensive air-showers, which can be observed by detector arrays deployed on the ground, which either directly collect secondary particles, or measure the \v{C}erenkov light or radio emissions of the air-shower particles. 

The geometrical optimization of such arrays requires to compute the air-shower signal received at each detection unit for various layouts. This computation can be performed with the pruning method as follows: 

\begin{enumerate}
\setlength\itemsep{-0.1em}
\item run a set of microscopic air-shower simulations on a layout optimized for interpolation,
\item define a dense array of detection units with regular spacing which will be the base array for all other layouts, 
\item interpolate the footprint of the simulations over the dense array, 
\item prune the information on unnecessary detection units to obtain the signal footprint over the desired layout. 
\end{enumerate}

For demonstration, we focus here on the radio air-shower detection technique. The radio technique is particularly well adapted to cover large areas, because radio antennas are cost-effective, robust and scalable. For inclined air-showers, produced by UHE particles interacting with the molecules of the atmosphere, the technique is all the more effective, as the beamed cone of the radio signal  propagates over several tens of kilometers of atmosphere, leaving a footprint on the ground that can reach several kilometers in diameter. This allows for an excellent sampling of the signal even with a sparse antenna array, hence enabling the deployment of detection units over gigantic areas. The AugerPrime Radio upgrade experiment \cite{Borodai_2022}, currently being deployed, targets these types of showers, as does the projected GRAND experiment \cite{GRAND20, Kotera:2021hbp, deMelloNeto:2023zvk} and its prototypes \cite{Zhang:2021tdh, Ma:2023siw}.

We will describe each of the steps of the pruning method tailored to the radio framework, in the following subsections.The numerical code containing steps 2.-4. and with example input simulations of 1., is publicly available\footnote{\href{https://github.com/kumikokotera/GRAND_tools}{https://github.com/kumikokotera/GRAND\_tools}}. 
   
\subsection{Microscopic simulations} \label{section:simulations}

The electromagnetic component of the air-showers generates an electromagnetic emission mainly through the deflection of charged particles by the geomagnetic field. This geomagnetic emission is coherent in the 10s of MHz frequency range, generating short ($< 1\,\mu$s), transient radio pulses, with high enough amplitudes for the detection of showers with energy $\gtrsim 10^{16.5}\,$eV \cite{Huege:2016veh,SCHRODER20171}. 

The ZHAireS numerical simulation \cite{Zhaires:2012} computes the electric field of air-showers by adding the contribution of every electron and positron at each step (track) in the shower. For every track in the simulation, there is one call to the field calculation routine per antenna, so the total CPU time is proportional to the number of required antennas $N_{\rm antenna}$. Inclined showers cover more distance in the atmosphere before arriving to ground level, requiring more tracks to be computed. Consequently, the total CPU time scales approximately as $N_{\rm antenna}/\cos\theta$,  were $\theta$ is the shower zenith angle, (0 for vertical, $90^\circ$ for horizontal). A single microscopic simulation with 160 antennas can take more than 60 CPU hours for very inclined showers. The interpolation of these simulations over dense arrays only take some seconds and their subsequent pruning allows to reuse them several times, which constitutes a substantial reduction in the CPU time required to analyse the performance of a given layout. 

For this study, we simulated a set of $N_{\rm sim} = 19200$ proton air-showers with ZHAireS, with observers positioned over a {\it star-shaped} pattern in the shower plane with eight arms, two aligned with the ${\bf v}\times{\bf B}$ axis, and two aligned with the ${\bf v}\times({\bf v}\times{\bf B})$-axis, where  ${\bf v}$ is the shower velocity along the axis and ${\bf B}$ the magnetic field of the Earth \cite{Buitink14}. 

The arrival directions of the simulated air-showers are parametrized by their zenith angle $\theta$ ($\theta=0^\circ$ corresponds to vertical air-showers). The zenith angles cover the range $[31^\circ-87^\circ]$ in bins following a uniform distribution in $\log_{10}(1/\cos\theta)$, as several key quantities in this study (size of footprint, distance to the radio emission point) scale with $1/\cos\theta$ ---and $1/\cos\theta$ covers a large range. The logarithmic distribution provides values of the zenith angle that result in footprints being 20\% bigger on each successive zenith bin. 
The energy of the protons initiating the air-showers is chosen among 20 energy bins, covering the range $\log_{10} (E/{\rm eV}) = 16.3-18.3$ with logarithmic steps of 0.1.

The simulations are run with a magnetic field of 56$\,\mu T$ and an inclination of 60.8\,deg, and the antennas are located at an altitude of 1086\,m above sea level, in the region of Dunhuang in North-Western China. Showers are simulating as coming from the North, South and the East to sample all the range of geomagnetic angles. 

Because the trigger criteria are intrinsically based on the voltage measured at the antenna level, it is necessary to infer the voltage information from our simulated electric fields. This requires to make an assumption on the antenna response of the detector.  

For illustrative purposes, we use in this study the antenna response of {\sc HorizonAntennas} described in Ref.~\cite{GRAND20}. The {\sc HorizonAntenna} is an active bow-tie antenna with a relatively flat response as a function of azimuthal direction and frequency. It has 3 perpendicular arms oriented along two horizontal directions and a vertical one. The {\sc HorizonAntenna} uses the same low-noise amplifier, but its radiating element is half the size of that in CODALEMA and AERA \cite{Charrier:2012zz,Abreu:2012pi}, in order to increase the sensitivity to the $50-200$\,MHz range. 

In the following and throughout this paper, the term `voltage' will refer to the measurement of the peak-to-peak voltage amplitude in the simulated signals, at each antenna.

\subsection{Layout geometries} \label{section:layouts}

Detection units can be placed at random locations, or following geometrical patterns with a unit cell repeated over a given surface. In this study, we focus on geometrical layouts with a hexagonal unit cell, which presents more symmetries compared to a rectangular grid structure. We also explore the effects of placing a denser array (an ``infill'') with various geometries inside a coarse array. 

\begin{figure}[tb]
\begin{center}
\includegraphics[height=0.4\textwidth]{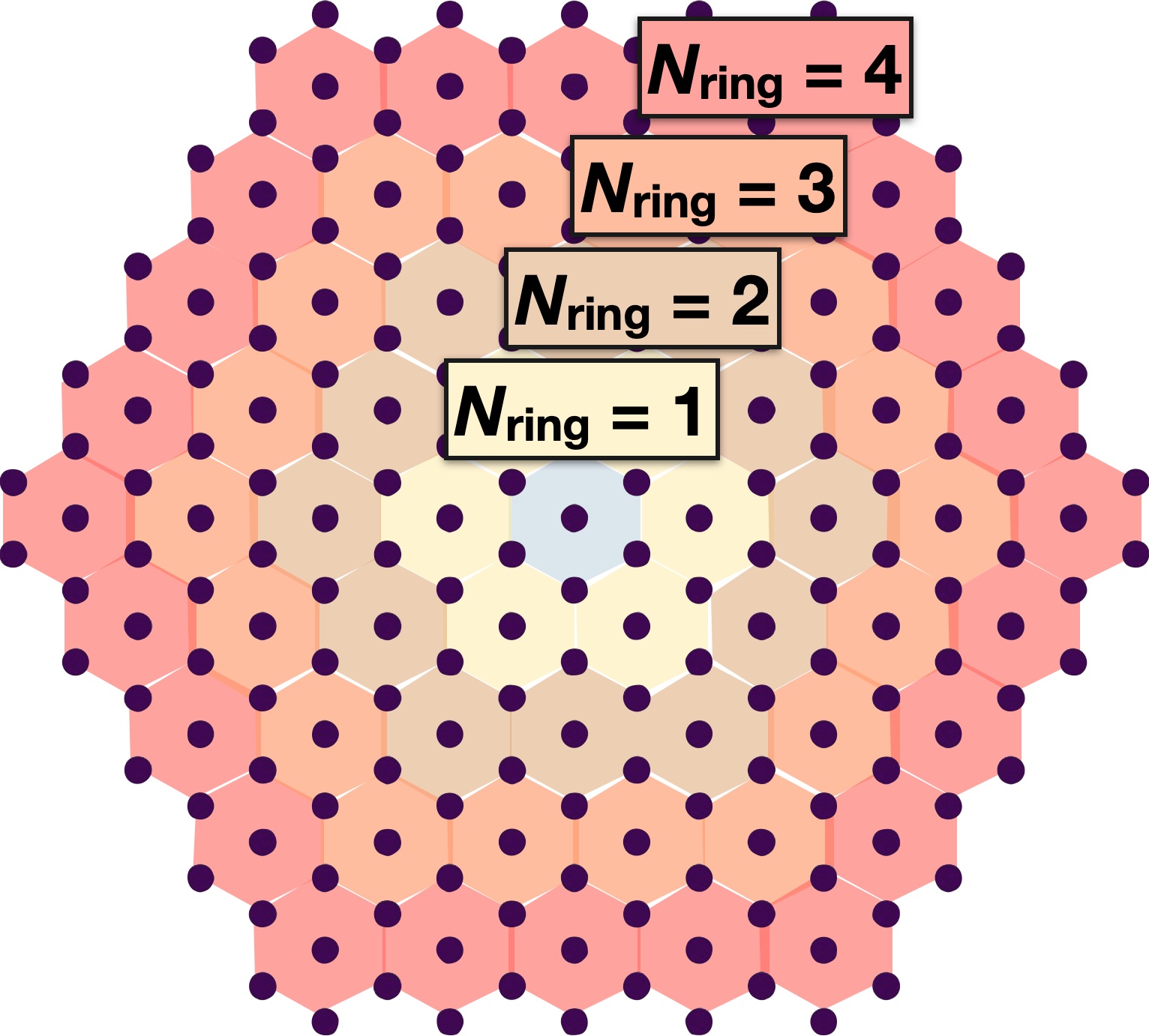} 
\includegraphics[height=0.4\textwidth]{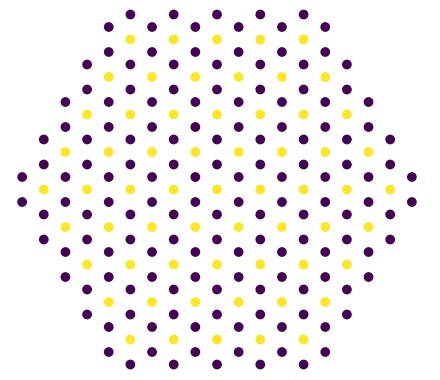} 
\caption{{\it Left:} Triangular `tri' array with $N_{\rm ring}=4$, with $N_{\rm ring}$ the numbers of concentric rings of hexagons around a central unitary hexagon, as indicated. The total number of antennas is $N_{\rm ant}=211$ {\it Right:} Hexagonal `hex' array with $N_{\rm ring}=4$ and $N_{\rm ant}=150$. The yellow points, corresponding to the centers of the hexagons, are \textit{pruned} from the 'tri' array to create the 'hex' array.}\label{fig:hex}
\end{center} 
\end{figure}

Our benchmark hexagonal grid corresponds to $N_{\rm ring}$ concentric rings of unitary hexagons around a central hexagon, which constitutes a first cell (see Fig.~\ref{fig:hex}, left). The step size of the array corresponds to the length of one side of the hexagon. Our routines to compare the performances of the arrays implement universal tools to manipulate hexagonal cells\footnote{\href{https://github.com/RedFT/Hexy}{https://github.com/RedFT/Hexy}}. We label these layouts `hex'. 
We also explore triangular grids (see Fig.~\ref{fig:hex}, right), which correspond to the hexagonal grids introduced above, appended with the central point of each hexagonal cell. Each hexagon is then split into 6 triangular cells.
The global shape of the array remains hexagonal. The step size of the array corresponds to the side of a triangular cell. We label these layouts `tri'.

\begin{figure}[tb]
\includegraphics[width=0.49\linewidth]{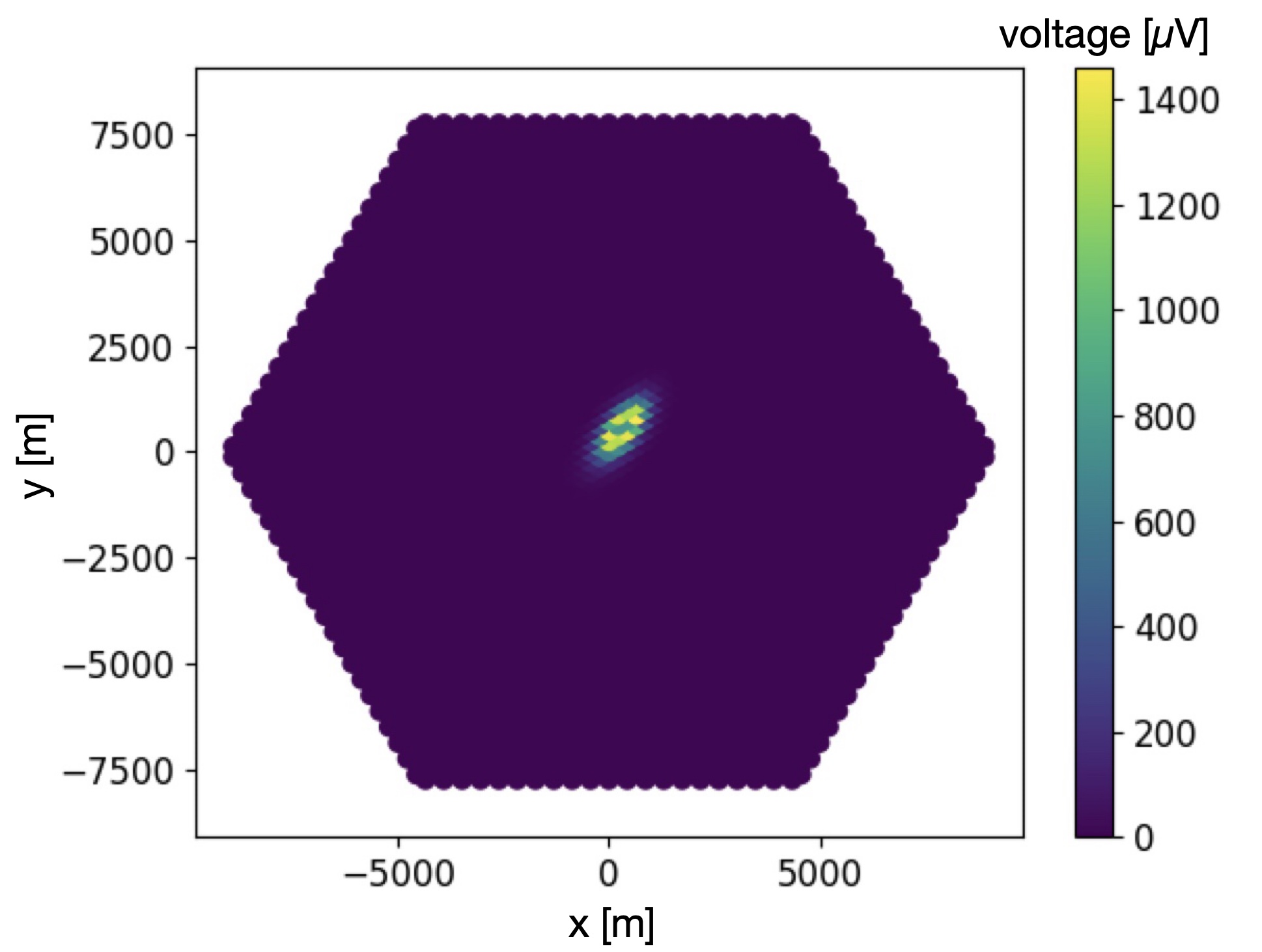}
\includegraphics[width=0.49\linewidth]{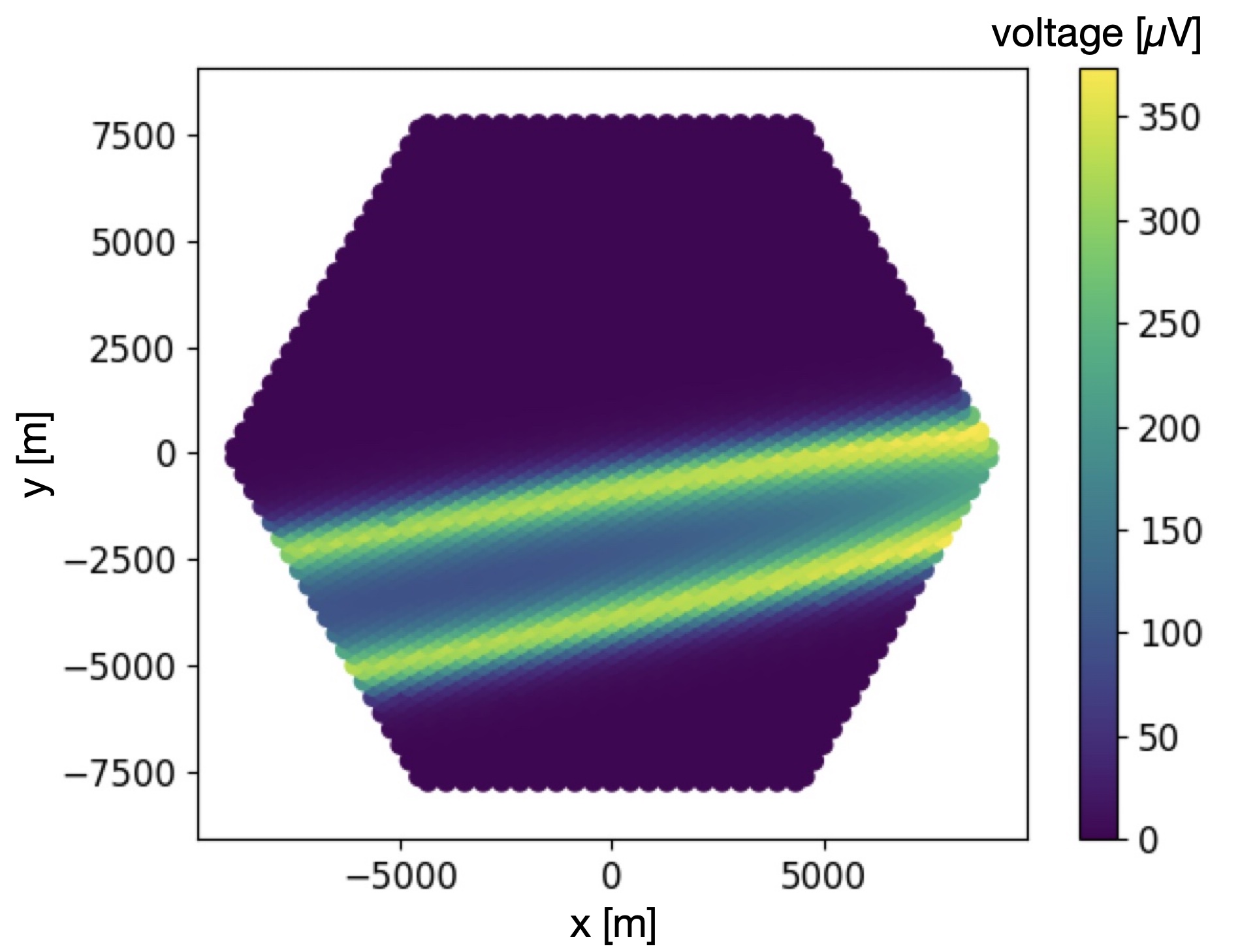}
\caption{Examples of footprints on the ground of the radio signal of air-showers, interpolated with the method of Ref.~\cite{Tueros21}, over a triangular array with an overall hexagonal shape (`tri'), with number of hexagon rings $N_{\rm ring}=20$. Each pixel represents the position of an antenna. The color code represents the voltage measured at each antenna in units of $\mu$V. Primary proton parameters: {\it Left:} energy $E=10^{18}\,$eV, zenith angle $\theta=62.98^\circ$ and {\it Right:} $E=10^{18}\,$eV,  $\theta=85,80^\circ$.}
\label{fig:footprints}
\end{figure}

In order to be able to cover a wide of possible layouts and step sizes we will use as a basis for all layouts explored in this article a 'tri' layout with $N_{\rm ring}=20$ with 250m step size, consisting of 3907 antennas, over which we will generate events by interpolation.

\subsection{Interpolation} \label{section:interpolation}

To generate events over the base layout, the simulations described in Section~\ref{section:simulations}, ran on star-shaped patterns, are interpolated over the layout geometries presented in Section~\ref{section:layouts} following the interpolation method of Ref.~\cite{Tueros21}. With that method, the expected full time traces of the signals can be synthesized at any point around the shower core from a set of signals on a star-shaped pattern, with  precision on the estimation of the signal amplitude of $-8/+15\%$ at a 95\% C.L., well below other sources of systematic errors in the simulation chain. Full event libraries can then be created over various geometrical layouts with a CPU time cost 3 orders of magnitude lower than that required to produce microscopic simulations (1h per interpolated event vs more than 1200h that would require a microscopic calculation over 3907 antennas). Furthermore, each microscopic simulation on the star-shape pattern can be re-utilized several times with different core positions to get as many events as needed.

Unless specified otherwise, the position of the shower core of each event is randomly shifted over the full array (the cores are {\it contained}). Strictly speaking, this implies that the event rates computed with these simulations are lower limits, as showers with cores landing outside of the array could also be detected. However, existing reconstruction methods are often based on a good reconstruction of the shower core, which requires to have active stations surrounding the core to better constrain its position. Using core-contained events thus offer a reasonable estimate of the number of events that could be reconstructed with good accuracy. After choosing a core position, a random rotation of the array is introduced to simulate a random incoming direction, to avoid artifacts in the results produced by the geometrical regularities present on the layout that wants to be tested.

Figure~\ref{fig:footprints} presents examples of interpolated voltages from simulations run over sets of antennas placed in a star-shaped pattern, into a triangular grid (`tri') of 3907 antennas (with $N_{\rm ring}=20$ concentric rings) forming a hexagon-shaped array. Each pixel represents and antenna, color-coded with the voltage level measured at each antenna, in units of $\mu$V. 

\begin{figure}[tb]
\includegraphics[width=0.49\linewidth]{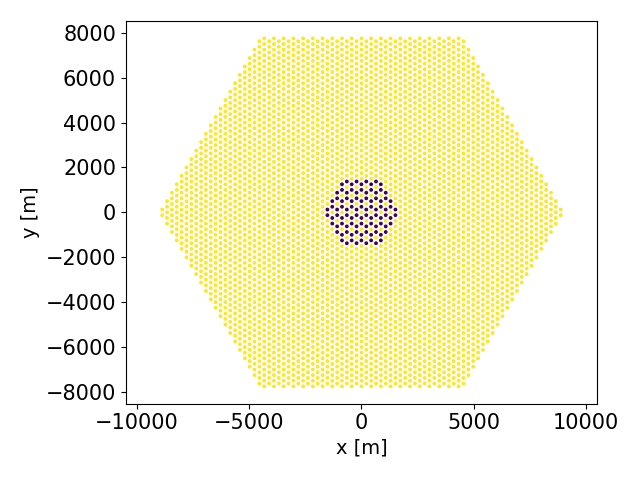}
\includegraphics[width=0.49\linewidth]{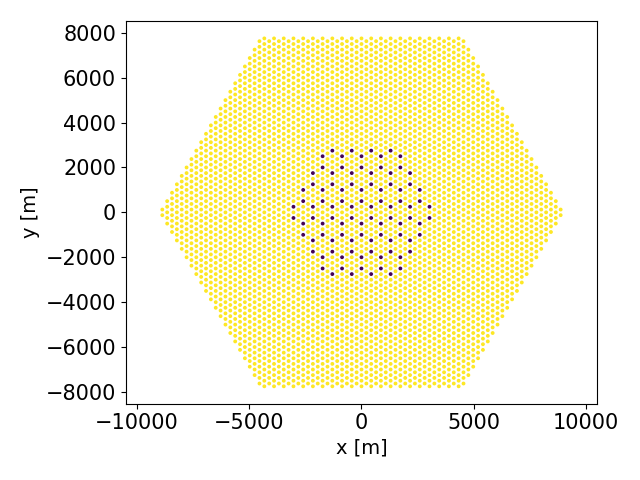}
\includegraphics[width=0.49\linewidth]{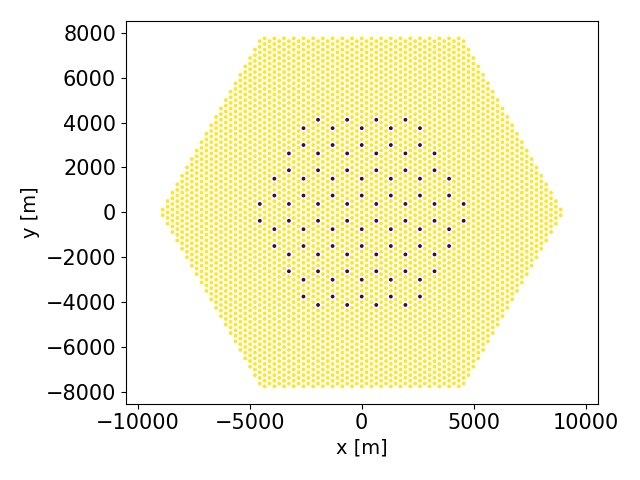}
\includegraphics[width=0.49\linewidth]{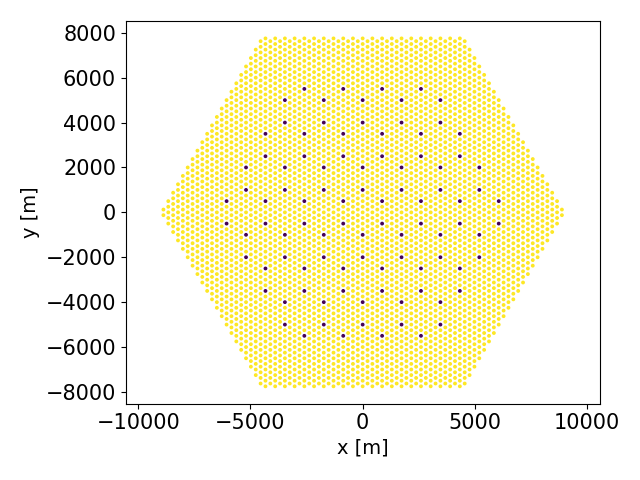}\\
\centering\includegraphics[width=0.49\linewidth]{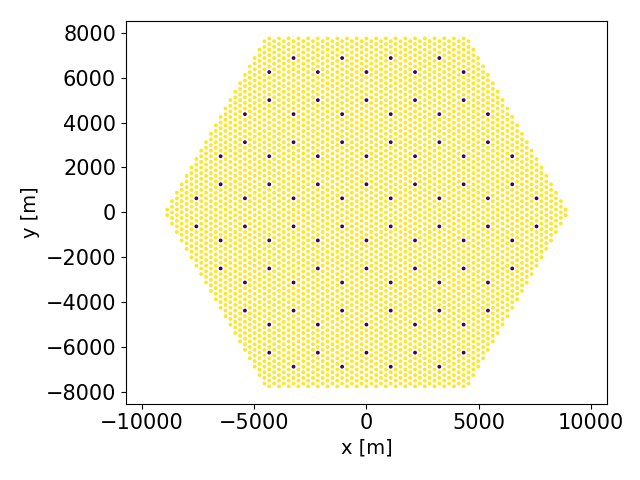}
\caption{Examples of hex antenna layouts (purple dots) pruned from a dense tri array with step size 250\,m (yellow pixels), with number of hexagon rings $N_{\rm ring} = 20$. All arrays contain a fixed number of $N_{\rm ant}=96$ antennas, with a step size of (from left to right, top to bottom): 250, 500, 750, 1000, 1250\,m.}
\label{fig:pruned_layouts}
\end{figure}

\subsection{Pruning}

Starting with a set of simulations/interpolations ran over a dense detector array, we prune selected detection units in order to obtain different layout shapes. For instance, by pruning the center of the hexagons of a tri array, one can obtain a hex array (left panel of Fig.~\ref{fig:hex}). Figure~\ref{fig:pruned_layouts} shows different levels of pruning starting from a dense tri array with $N_{\rm ring}=20$ and step size $250\,$m. 

Pruning gives us the ability to quickly test any array shape that can be synthesised with the original dense array, including infills, irregular arrays or even random arrays (see Fig.~\ref{fig:infill_layouts} for some examples), using the same simulation library, and with almost no additional CPU time required (pruning an event takes less than a second). 

In practice, the pruned layout is generated from a dense array as follows. A new sparser array is produced, as a list of coordinates of the detector positions. These positions are compared to those in the initial dense array. Only the signal information of the matching detection units in the initial dense array are kept.

\section{Exploring layouts with the pruning method}\label{section:exploring}

The choice of a specific array layout is intimately linked to the targeted science case.
Layouts are chosen to optimize the detection rate (optimize the triggering efficiency) as well as the reconstruction performances for primary particles with targeted natures, energies, arrival directions, and possibly other properties. 

In order to assess the performances of a given layout, three sets of parameter spaces are to be explored:
\begin{enumerate}
\item Parameters related to primary particles. In this study, we will focus in particular on the primary particle energy $E$ and zenith angle of the arrival direction $\theta$. Depending on the science case, one will target a specific region in this parameter space, or to broaden the range maximally. 
\item Parameters related to layout geometries. These are typically the step size $\lambda$, total number of antennas $N_{\rm ant}$, geometry, and for non homogeneous layouts: granularity. These are the parameters on which we have leverage, though taking into account constraints in terms of total number of antennas (budget), total area covered (available land), spare antennas available for a possible infill etc. 
\item Parameters related to detection performances. These correspond for example to parameters related to triggering criteria: the trigger threshold in terms of signal (e.g., voltage $V_{\rm trig}$), and the number of antennas required for trigger $N_{\rm trig}$. Quality cuts can also be introduced for optimized reconstruction. Note that these parameters are usually quite uncertain until the experiment is actually operational, but ranges or lower limits are available, enabling to explore conservative or optimistic scenarios. 
\end{enumerate}

Depending on the science case, one will want to improve the sensitivity in various regions of the parameter space (1). In the following, we will make use of the pruning method to explore the effects of various layout parameters (2). We do not specify a science case, and do not adopt refined quality cuts for reconstruction. For the parameter space (3), we use simple triggering conditions and use core-contained events, i.e., air-showers for which the center (or core) lands within the array. This latter criterion enables to introduce a basic quality cut for reconstruction, which is usually facilitated when the core position is identified. 

In our exploration, we will adopt as layout quality estimators the triggering efficiency , the event detection rates and heat maps when assessing the sensitivity of infills. In the general case, once a specific science case and clear objectives have been drawn, one can optimize the layout taking these effects into account, using for instance bayesian optimization techniques. 

\subsection{Triggering efficiency and effective area} \label{section:trig_eff}

An event is defined as triggered if $N\ge N_{\rm thres}$ antennas detect a voltage $V\ge V_{\rm thres}$. We note $N_{\rm trig}(E,\theta)$ the total number of triggered events at energy $E$ and zenith angle $\theta$. The trigger efficiency at $(E,\theta)$ is then estimated as $r_{\rm trig}(E,\theta)=N_{\rm trig}(E,\theta)/{N_{\rm tot}(E,\theta)}$, where ${N_{\rm tot}(E,\theta)}$ is the total number of showers simulated at $(E, \theta)$. 

The effective area $A_{\rm eff}$ is then defined as the area of a simulation set, weighted by the triggering efficiency
\begin{equation}
A_{\rm eff}(E,\theta) = r_{\rm trig}(E,\theta)\, A_{\rm sim}\cos\theta\ , 
\end{equation}
where $A_{\rm sim}\cos\theta$ is the surface of the full simulated array orthogonal to the shower axis\footnote{Fluxes are defined as $\Phi = \int {\rm d} {\bf S}\cdot {\bf F}$, hence the relevant surface for flux or event number calculations corresponds to the surface projected to the plane orthogonal to the air-shower axis.}.

\subsection{Differential event detection rates}\label{section:diff_ev_rate}
The event detection rate (number of events day$^{-1}$\,km$^{-2}$\,sr$^{-1}$\,PeV$^{-1}$) over the full array can be estimated by folding in the cosmic ray flux $J_{\rm cr}$ measured by the TALE experiment \cite{2018ApJ...865...74A} to the trigger efficiency and effective areas calculated in the previous section:
\begin{equation}\label{eq:nu_ev}
\frac{{\rm d}N_{\rm ev}}{{\rm d}t\,{\rm d}\Omega\,{\rm d}E} (E,\theta) = r_{\rm trig}(E,\theta)\, A_{\rm sim}\cos\theta\,J_{\rm cr}(E) \ ,
\end{equation}
where ${\rm d} \Omega=2\pi \sin \theta {\rm d} \theta$ the solid angle element. The total number of events detected by the array, over a given time frame $\Delta t$, energy range $\Delta E$ and solid angle $\Delta\Omega$ is calculated by integrating Eq.~(\ref{eq:nu_ev}).

\subsection{An ideal dense array to provide insight on the energy/inclination interplay for radio footprints}

\begin{figure}[tb]
\includegraphics[width=0.49\linewidth]{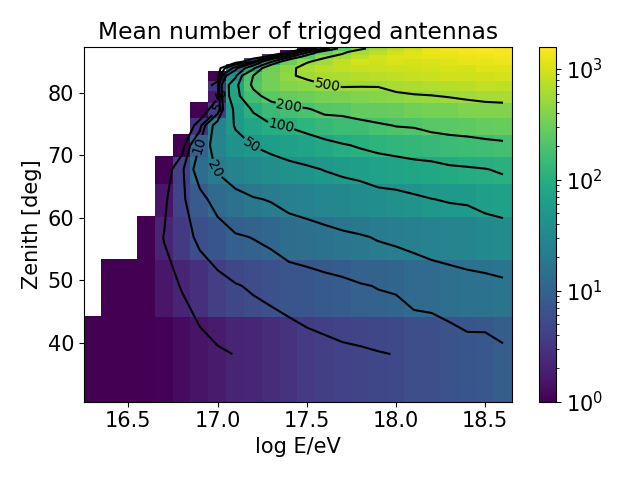}
\includegraphics[width=0.49\linewidth]{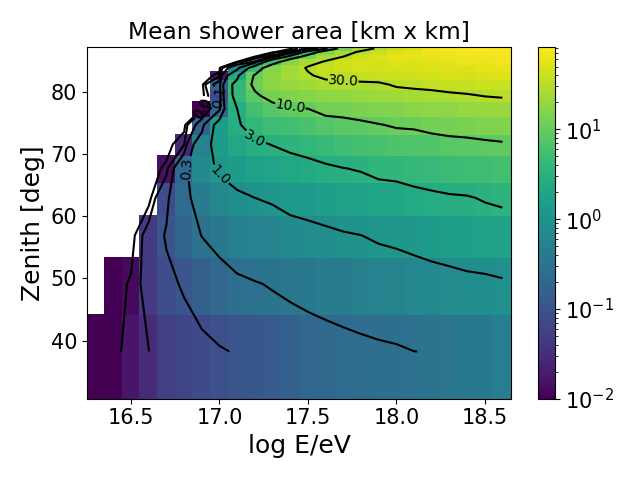}\\
\centering \includegraphics[width=0.49\linewidth]{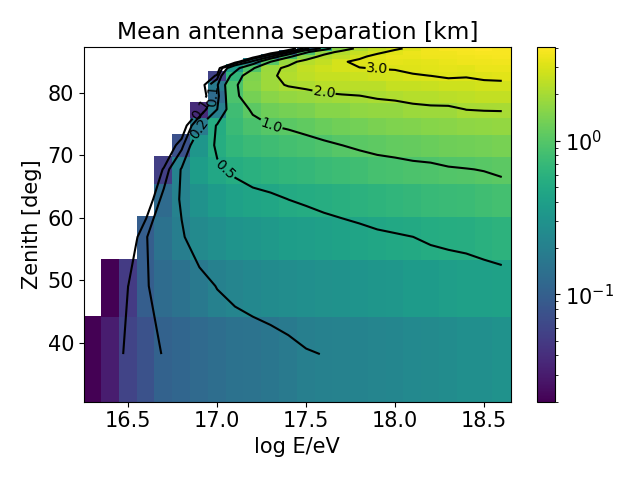}
\caption{Rule-of-thumb exploration of the interplay between the energy and arrival directions on the area of the radio footprint with signal above $V_{\rm trig}$, and hence on the required array step size. Quantities are calculated on a dense hexagonal array with step size $\lambda=250\,$m ($N_{\rm ring}=20$) and surface $A_{\rm sim}\sim 200\,{\rm km}^2$, with a total of $N_{\rm ant}=3907$ antennas. {\it Top left}: mean number of triggered antennas per event $\bar{n}_{\rm trig}$ (color contours) as a function of shower zenith angle $\theta$ and primary energy $E$. {\it Top right}: estimate of the area of the footprint $a_{\rm footprint}$, as in Eq.~(\ref{eq:afoot}), in the $(\theta,E)$ parameter-space. {\it Bottom}: mean antenna separation as in Eq.~(\ref{eq:mean_sep}) in the $(\theta,E)$ parameter-space.}
\label{fig:mean_n_trig}
\end{figure}

We first consider a dense array hexagonal array with step size $\lambda=250\,$m ($N_{\rm ring}=20$) and surface $A_{\rm sim}\sim 200\,{\rm km}^2$, with a total of $N_{\rm ant}=3907$ antennas. This is not necessarily a realistic array, but its fine sampling and large surface enables us to probe general effects on the footprints, related to physical parameter space (1) invoked above, without taking into consideration any layout parameters~(2). Sparser arrays will be considered in the next sections. All simulated events have their shower core position inside the array. An antenna is defined as triggered if it measures a voltage higher than $V_{\rm trig}=75\,\mu$V, which is considered to be 5 times the expected galactic noise. A shower event is defined as triggered if more than $N_{\rm trig}=5$ antennas are triggered.

In Figure~\ref{fig:mean_n_trig}, we explore the interplay between the energy and the arrival inclination of the showers (parameters 1) on the size of the footprint, which can guide the choice of a specific array layout. The top left-hand contours present the mean number of triggered antennas per event $\bar{n}_{\rm trig}$ as a function of zenith angle and primary energy. 
For a given simulation area $A_{\rm sim}$  
and a given total number of antennas $N_{\rm ant}$, 
the mean area where the radio signal from a shower leads to a voltage at the antennas above the triggering threshold can be estimated as:
\begin{equation}\label{eq:afoot}
A_{\rm footprint}(E,\theta) = \frac{A_{\rm sim}}{N_{\rm ant}}\bar{n}_{\rm trig}(E,\theta) \ .
\end{equation}
This quantity, which can be used as an estimate of the area of the radio footprint with signal above $V_{\rm trig}$, is represented in the top right-hand panel of Figure~\ref{fig:mean_n_trig}. 

By dividing this mean footprint area $a_{\rm footprint}$ by the number of antenna needed to trigger $N_{\rm trig}$, and taking the square root of this quantity, we obtain what we call the mean antenna separation:
\begin{equation}\label{eq:mean_sep}
\bar{\lambda}(E,\theta) = \sqrt{\frac{a_{\rm footprint}(E,\theta)}{N_{\rm trig}}} \ .
\end{equation}
This quantity corresponds to an order-of-magnitude estimate of the separation between antennas needed to trigger events as a function of zenith angle and primary energy. It is represented in the bottom-left-hand panel of Figure~\ref{fig:mean_n_trig}. For example, an event with arrival direction $\theta=60^\circ$ and energy $E = 10^{17.75}$\,eV can be triggered by considering a separation of about 500\,m. 

Figure~\ref{fig:mean_n_trig} supports the intuition that for large zenith angles and high particle energies, footprints are large and sparse arrays are acceptable. On the other hand, for low zenith angles and particles with low energies, denser arrays are required to enable detection and for good signal sampling.

\subsection{Geometry effects}\label{section:geom}

We compare the effect of adding a central detector in each unitary hexagonal cell, i.e., going from `hex' to `tri' geometries, keeping the same step size. Table~\ref{tab:pruning_layout} lists the number of antennas required for each combination of geometry and step size to instrument a surface of $A_{\rm sim}=204\,$km$^2$. 

The last two columns of the table provides the number of detected events on the array per day $N_{\rm ev}$ in our computations, as well as this quantity divided by the total number of antennas $N_{\rm ev}/N_{\rm ant}$. $N_{\rm ev}$ is calculated following Section~\ref{section:diff_ev_rate}. These numbers are purely indicative, meant for rough comparison purposes only. In particular, we do not quote here the statistical uncertainties due to the limited number of simulations run in each energy and zenith bins. 

It is however interesting to note that in all the geometries and step sizes explored in Table~\ref{tab:pruning_layout}, each antenna  detects between 1 and 2 events per day on average. 

\begin{figure}[tb]
\includegraphics[width=0.49\linewidth]{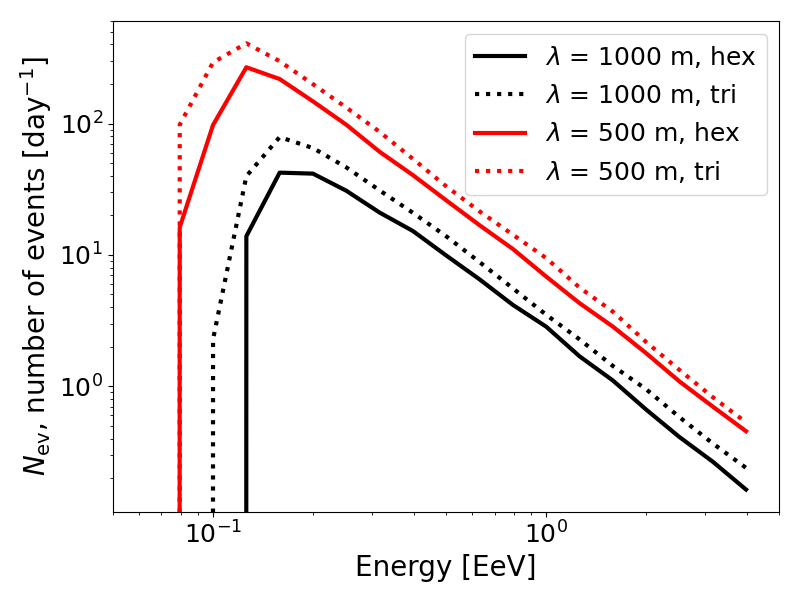}
\includegraphics[width=0.49\linewidth]{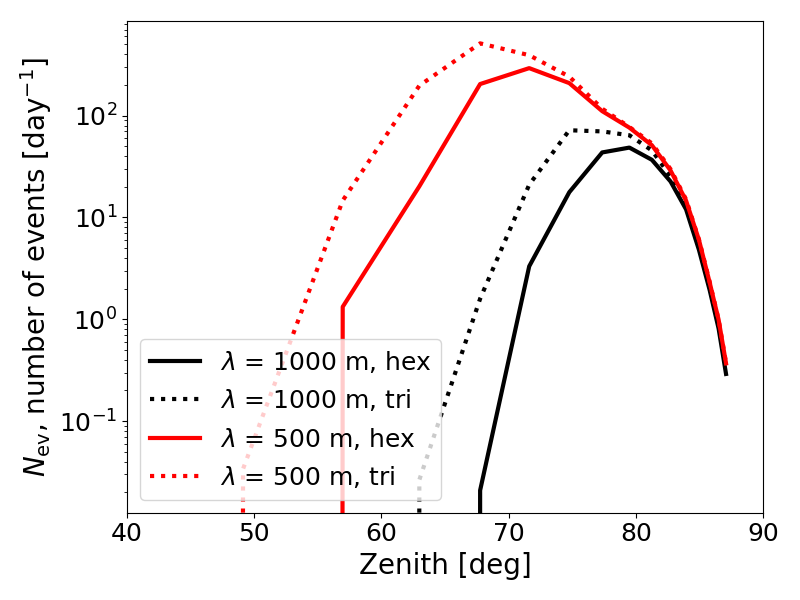}
\caption{Influence of geometrical pattern of the layout on the detection rates. Comparison of detection rates for `hex' and `tri' layouts for a fixed surface of $A_{\rm sim}=204\,$km$^2$. Detection rates computed for $V_{\rm trig}=75\,\mu$V and $N_{\rm trig}=5\,$antennas, for $\lambda=1000\,$m step (black) and $\lambda=500\,$m step (red). The total number of antennas for `hex' layouts (solid lines) and  `tri' layouts (dotted lines) are indicated in Table~\ref{tab:pruning_layout}.  }
\label{fig:hex_vs_tri}
\end{figure}

\begin{figure}[tb]
 \centering
\includegraphics[width=0.49\linewidth]{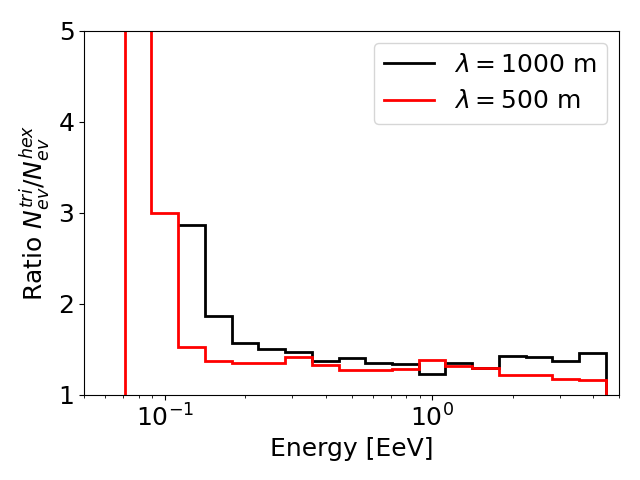}
\caption{Ratio of number of detected events $N_{\rm ev}$ per day for `hex' and `tri' layouts for a fixed surface of $A_{\rm sim}=204\,$km$^2$. Detection rates computed for $V_{\rm trig}=75\,\mu$V and $N_{\rm trig}=5\,$antennas, for $\lambda=1000\,$m step (black) and $\lambda=500\,$m step (red). The total number of antennas for `hex' layouts (solid lines) and  `tri' layouts (dotted lines) are indicated in Table~\ref{tab:pruning_layout}.}
\label{fig:hex_vs_tri_ratio}
\end{figure}

In terms of number of antennas, one can notice that the `tri' geometries require typically $50-60\%$ more antennas to instrument the same area. Figure~\ref{fig:hex_vs_tri} shows a moderate gain in the detection of low energy particles (left-hand panel), corresponding to more vertical events with lower zenith angles (right-hand panel), for triangular geometries.
However, Fig.~\ref{fig:hex_vs_tri_ratio} and the last two columns of Table~\ref{tab:pruning_layout} demonstrate that the overall gain in the total number of detected particles is marginal for $\lambda=500\,$m ($\lesssim 30\%$), especially at high energies. For a sparser array of $\lambda=1000\,$m, the gain can reach $\gtrsim40\%$ but does not compensate for the additional $60\%$ antennas to be deployed. 

On the other hand, at the low energy extremity, the `tri' layout presents a drastically better rate. The sharp increase at low energies observed in Fig.~\ref{fig:hex_vs_tri_ratio} can be explained by the fact that at low energies, footprints on the ground are small enough to be almost never triggering for a `hex' array, whereas a `tri' array can trigger thanks to the additional antenna in the center of every hexagon.

We propose in Section~\ref{section:infill} that excess antennas can have an efficient usage in a denser infill configuration, to increase the detection efficiency for low energy and less inclined events. A coarse hexagonal array of `hex' type seems to be sufficient and efficient to instrument a given area and obtain a significant sensitivity with a reasonable number of antennas. 

\begin{table}
 \centering
 \begin{tabular}{lcrrr}
   \hline \hline
   geometry & step size & $N_{\rm antenna}$ &$N_{\rm ev}$& $N_{\rm ev}/N_{\rm antenna}$\\
   &  $\lambda$ [m] &  & [day$^{-1}$] &  [day$^{-1}$]\\
   \hline 
   tri & 1000 &  241 & 321.77 & 1.33\\
   tri & 500 &  991 & 1660.68 & 1.67\\
   hex & 1250 & 96 & 115.04 & 1.20\\
   hex & 1000 &  150 &192.38 & 1.28\\
   hex & 750 & 294 &379.53 & 1.29\\
   hex & 500 &  660 &1018.72 & 1.54\\
   hex & 250 & 2646& 4570.05 & 1.72\\
   \hline
   \end{tabular}
   \caption{Characteristics of the 'tri' and 'hex' layouts covering a fixed surface of $A_{\rm sim}=204\,$km$^2$, with given step size $\lambda$.
   The layouts are obtained by pruning an initial dense tri layout with $N_{\rm ring}=20$ and step size 250\,m. For illustrative purposes, the last two columns provide an indication of the total number of events detected per day, and the ratio between this quantity and the total number of antennas, for each layout (see Section~\ref{section:geom} for cautionary note on these numbers).}
  \label{tab:pruning_layout}
\end{table}

\subsection{Step size effects}

\begin{figure}[tb]
\includegraphics[width=0.49\linewidth]{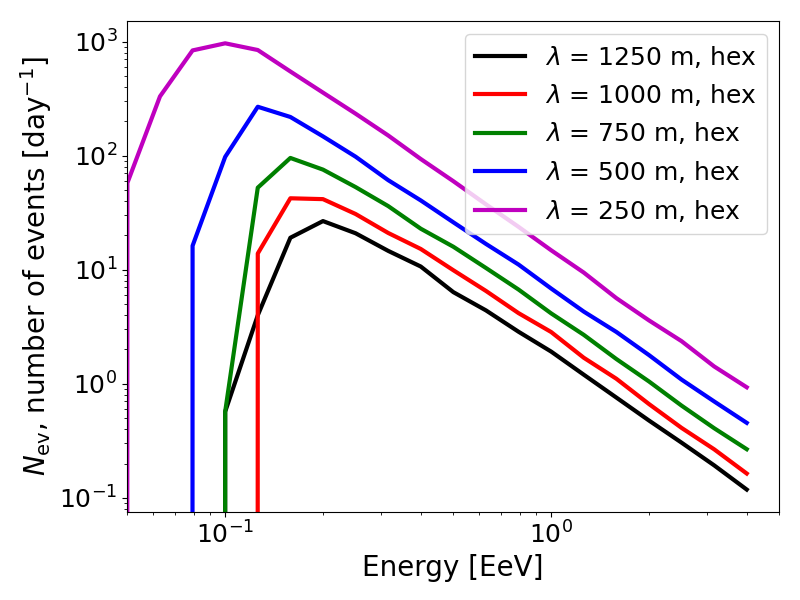}
\includegraphics[width=0.49\linewidth]{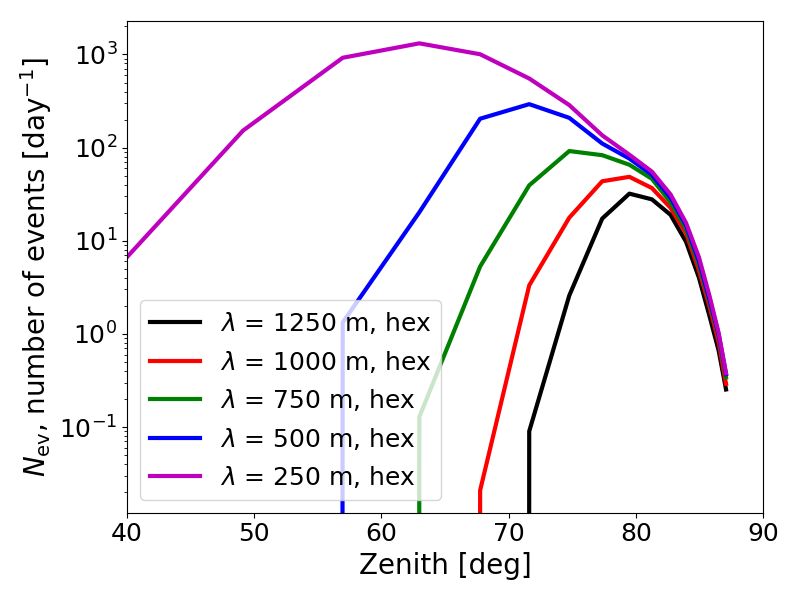}
\caption{Influence of antenna separation (or step size) on the detection rates. Comparison of detection rates for 'hex' layouts with step sizes $\lambda=250, 500, 750, 1000, 1250\,$m for a fixed surface of $A_{\rm sim}=204\,$km$^2$. Detection rates computed for $V_{\rm trig}=75\,\mu$V and $N_{\rm trig}=5\,$antennas. The corresponding total number of antennas are indicated in Table~\ref{tab:pruning_layout}.  }
\label{fig:steps}
\end{figure}

In order to evaluate the effects of different step sizes, we compute the detection rates for `hex' arrays of fixed surface $A_{\rm sim}=204\,$km$^2$. The results are presented in Fig.~\ref{fig:steps}. The large gain at low energies and small zenith angles appears clearly for dense arrays of step $\lambda=250\,$m. The dependencies on the step sizes can be quantitatively understood from the bottom panel of Fig.~\ref{fig:mean_n_trig}. Mean antenna separations $\lambda <500\,$m enable to cover a significant portion of the $(E,\theta)$ parameter-space. 
This effect is clearly visible in Fig.~\ref{fig:steps}. The gain in event rate at low energy and small zenith angles fully compensates for the larger number of antennas deployed (see also the last column of Table~\ref{tab:pruning_layout}). Choosing an array with step size $\lambda=750, 1000,$ or 1250\,m does not make a significant difference on the total event detection rate, or on the science case which can be explored. In order to enhance the sensitivity to lower energy and hence more vertical air-showers, it is more useful to instrument a specific denser region and create an infill array, as discussed in Section~\ref{section:infill}.

\subsection{Testing triggering criteria}

\begin{figure}[tb]
\includegraphics[width=0.49\linewidth]{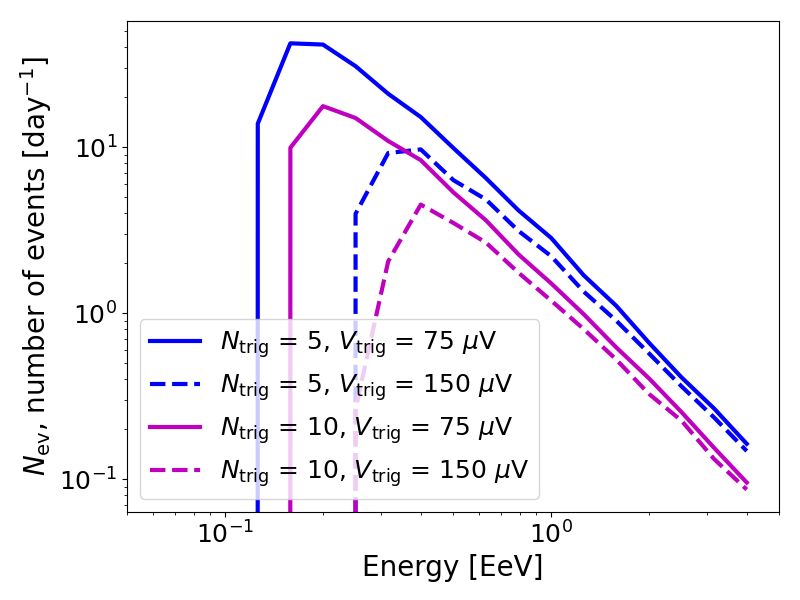}
\includegraphics[width=0.49\linewidth]{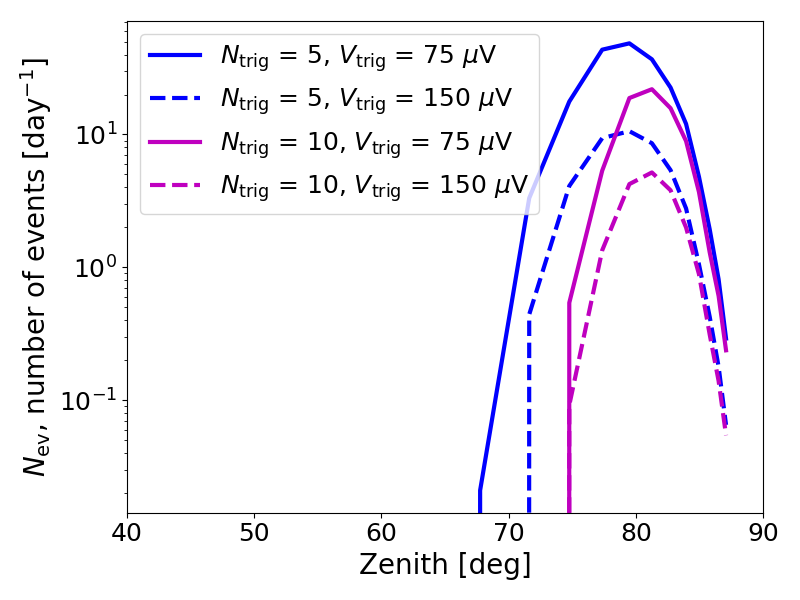}
\caption{Influence of simple triggering critera on the detection rates. Comparison of detection rates for 'hex' layouts with step sizes $\lambda=1000\,$m for a fixed surface of $A_{\rm sim}=204\,$km$^2$, for $V_{\rm trig}=75\,\mu$V (solid) and $V_{\rm trig}=150\,\mu$V (dashed) and $N_{\rm trig}=5$ (blue) and $N_{\rm trig}=10$ (magenta) antennas. The total number of antennas is $N_{\rm ant} = 150$.  }
\label{fig:trig_comp}
\end{figure}

The pruning method also enables us to easily test various simple triggering criteria. Once the array has been pruned to a new layout, various triggering criteria can be applied as a second step, to calculate detection efficiencies. 

Here, we illustrate this feature by changing two triggering threshold parameters: the minimum voltage measured at the antenna $V_{\rm trig}$ to trigger the acquisition, and the minimum number of antennas $N_{\rm trig}$ to experience such a level of voltage coincidentally. 

Our results are presented in Fig.~\ref{fig:trig_comp}. 
The voltage level naturally impacts the energy threshold of the detection rate, as higher shower energies translates into a brighter radio emission, hence an increased detectability. The minimum number of antennas has little influence on the low energy threshold, as the footprints above $0.1\,$EeV are larger than the surface covered by 10 antennas for zenith angles $\gtrsim 75^\circ$. The total number of events at peak energy is however lower of a factor of a few for $N_{\rm trig}=10$ compared to 5 antennas. 

The drastic cut on the zenith angles, allowing only for the detection of very inclined events with $\theta\gtrsim 75^\circ$ can be moderately compensated by relaxing the criterion on the minimum number of antennas triggered $N_{\rm trig}$, probably at the expense of reconstruction accuracy. If $N_{\rm trig}$ needs to be kept constant, the only solution would be to use a smaller step size, i.e., an infill.

\subsection{An infill to extend the sensitivity to low energies and smaller zenith angles}\label{section:infill}

One of the advantages of the pruning tool is to be able to explore various intricate structures, in particular for a dense infill array. We illustrate this process with few examples in this section. 

\begin{figure}[tb]
\includegraphics[width=0.49\linewidth]{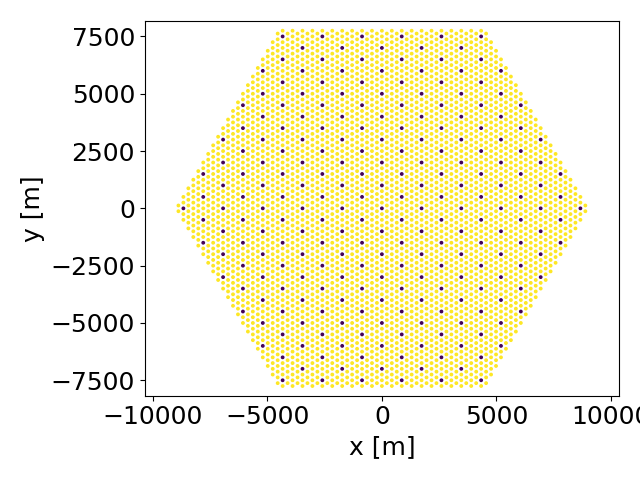}
\includegraphics[width=0.49\linewidth]{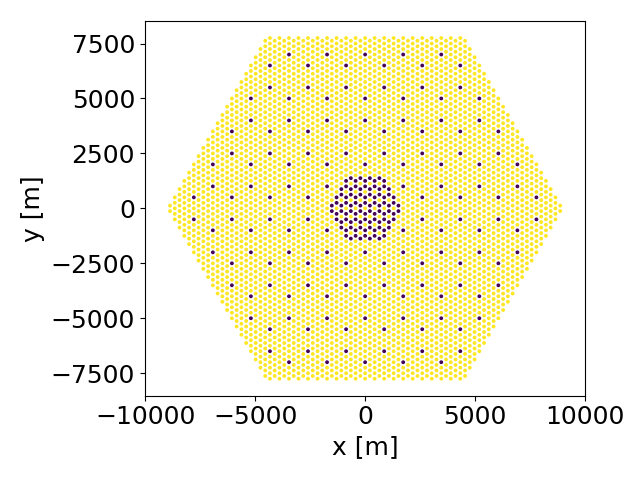}
\includegraphics[width=0.49\linewidth]{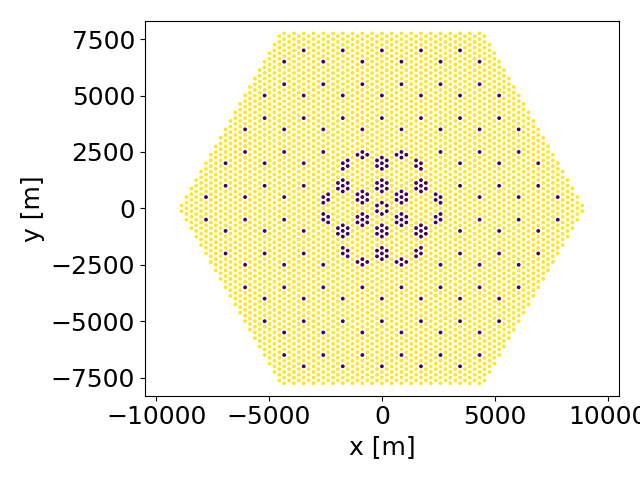}
\includegraphics[width=0.49\linewidth]{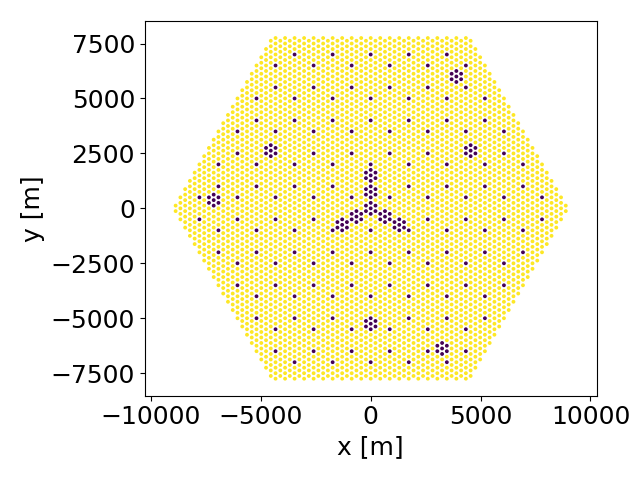}
\caption{Examples of infill antenna positions, on top of a coarse `hex' antenna layout with $\lambda = 1000\,$m and fixed surface $204\,$km$^2$ (purple dots), pruned from a dense tri array with step size 250\,m (yellow pixels). The array characteristics are documented in Table~\ref{tab:infill}. From left to right, top to bottom, the layouts are labeled: `tri 1000', 'island', 'flower' and 'spiral'.}
\label{fig:infill_layouts}
\end{figure}

Starting as usual from the dense array with step size $\lambda = 250\,$m, we construct various infill geometries, which are presented in Fig.~\ref{fig:infill_layouts}, with their corresponding characteristics listed in Table~\ref{tab:infill}. The infills are added to a coarse `hex' layout of step size $\lambda=1000\,$m. All infill layouts have proximatelly the same number of antennas. 

The most classical way to add an infill is to simply add one point at the center of each hexagon pattern. This way, we obtain our previously studied `tri' geometry, with step size $\lambda = 1000\,$m (top left panel). Another straighforward option is to add a dense regular sub-array (or island) in the center, keeping the hexagonal geometry (top right panel). We call this layout `island'. 

A more sophisticated approach could be to replace each of the hexagon extremities in the coarse array by dense mini-islands (more precisely here: a `tri' structure with $N_{\rm ring} = 1$). This is the concept of the `flower' infill pattern, which proposes a trade-off to enable the detection of lower energy and more vertical showers thanks to the islands, but instrumenting a larger infill surface than in the `island' case. Here the pattern chosen corresponds to a truncated 2-uniform tilings of the Euclidean plane, with vertex type $[3^6; 3^2.6^2]$\footnote{\href{https://en.wikipedia.org/wiki/Euclidean\_tilings\_by\_convex\_regular\_polygons}{Wikipedia page on Euclidean tilings by convex regular polygons.}}.

A final alternative that we explore is to tile the layout with such mini-islands, but in a logarithmic spiral structure over the full array. This should enable to smooth the transition towards more vertical/lower energy showers, thanks to the gradual spacing between the mini-islands. 
The pattern in the spiral layout comprises three spirals. The first spiral follows the following expression in polar coordinates: $r = a b^\theta$. The other two spirals are obtained by rotation by an angle $2\pi/3$ of the initial spiral. Positions along those spirals are then sampled and mini-islands are added to the coarse array. For the layout presented here, we have chosen $a=1000\;\rm{m}$, and $b=1.8$. 

\begin{figure}[tb]
\includegraphics[width=0.49\linewidth]{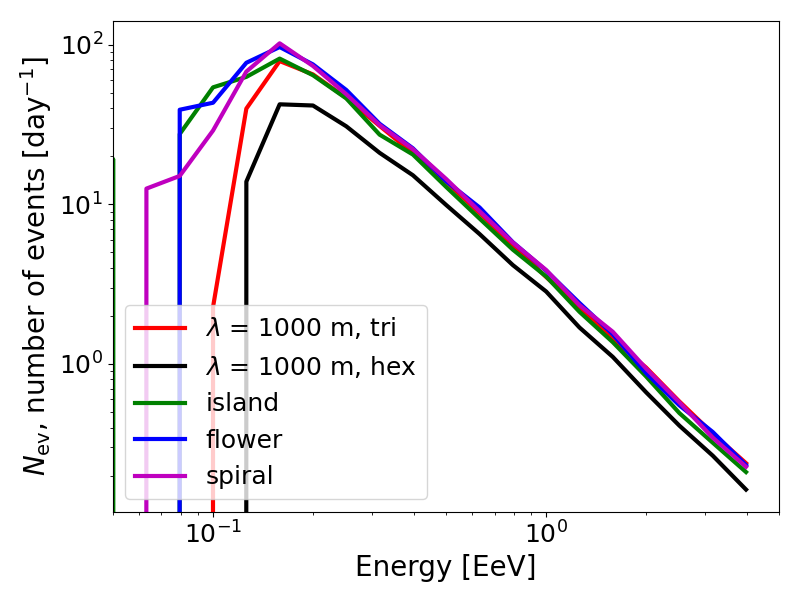}
\includegraphics[width=0.49\linewidth]{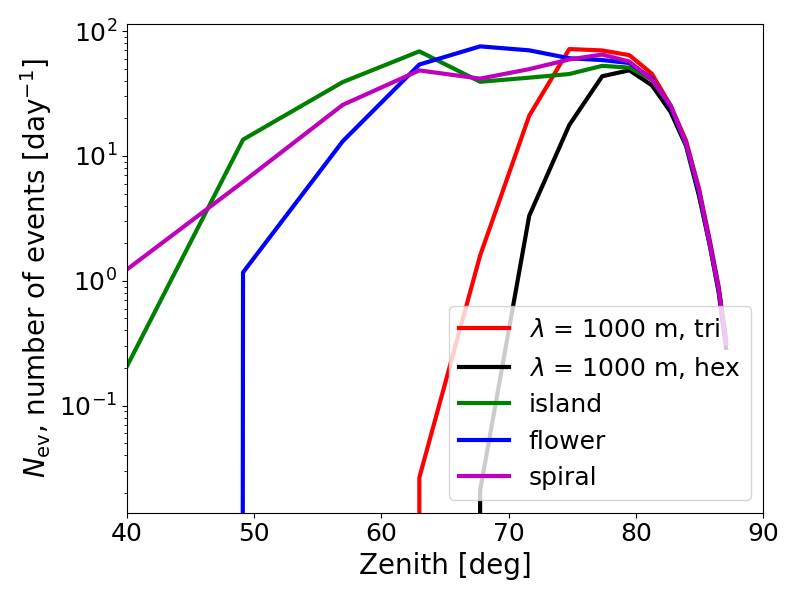}
\caption{Influence on the detection rates of infill addition to a coarse `hex' array, of step size $\lambda=1000\,$m for a fixed surface of $A_{\rm sim}=204\,$km$^2$, for $V_{\rm trig}=75\,\mu$V. 
The array characteristics are documented in Table~\ref{tab:infill}, and their patterns are presented in Fig.~\ref{fig:infill_layouts}. }
\label{fig:infill}
\end{figure}

In Figure~\ref{fig:infill}, we compute the detection rates for the four infill layouts, which all contain the same number of antennas. It strikingly appears that a regular infill with no granularity (the `tri' array obtained by putting an antenna at the center of each exagon) is not beneficial to extend the detection rates at low energy and small zenith angles. Spare antennas are better employed in granular infills. The island and flower patterns reach similar efficiencies at low energies. The island layout is best suited to achieve low zenith angles, but does slightly less sensitivity in the $\theta =65-75^\circ$ range, compared to the flower infill. 
The spiral infill, as expected, enables to achieve detection over a larger range in energy and zenith angles, but at the cost of an overall lesser sensitivity below the peaks. 

\begin{figure}[tb]
\includegraphics[width=\linewidth]{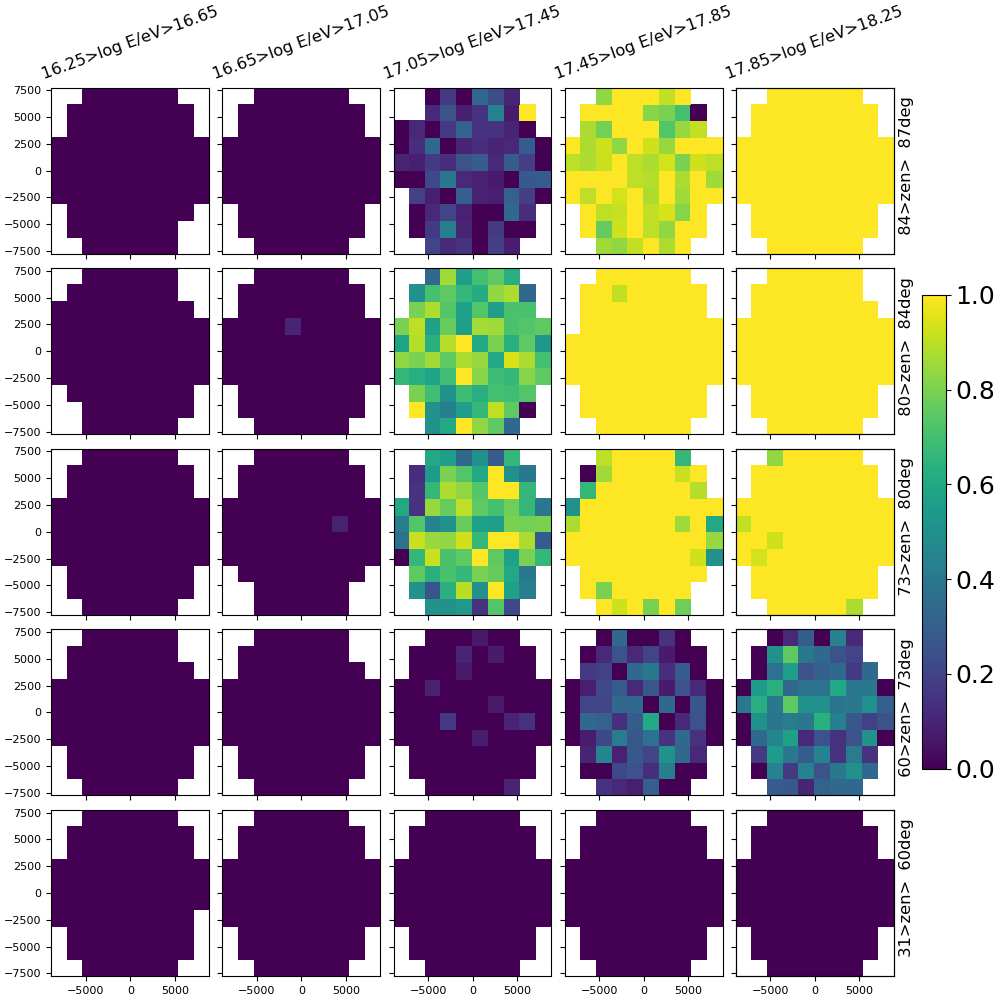}
\caption{Heat map for the `tri 1000' layout. Each panel  indicates the trigger efficiency $r_{\rm trig}$ (see Section~\ref{section:trig_eff}), computed in each pixels of size $x\times y =2.75\,{\rm km}^2$. Each panel-column corresponds to simulated showers in a range in energy $E$ as indicated in the upper labels. Each panel-row corresponds to showers a range in zenith angles $\theta$ as indicated in the right-hand labels.}
\label{fig:heatmap_tri}
\end{figure}

\begin{figure}[tb]
\includegraphics[width=\linewidth]{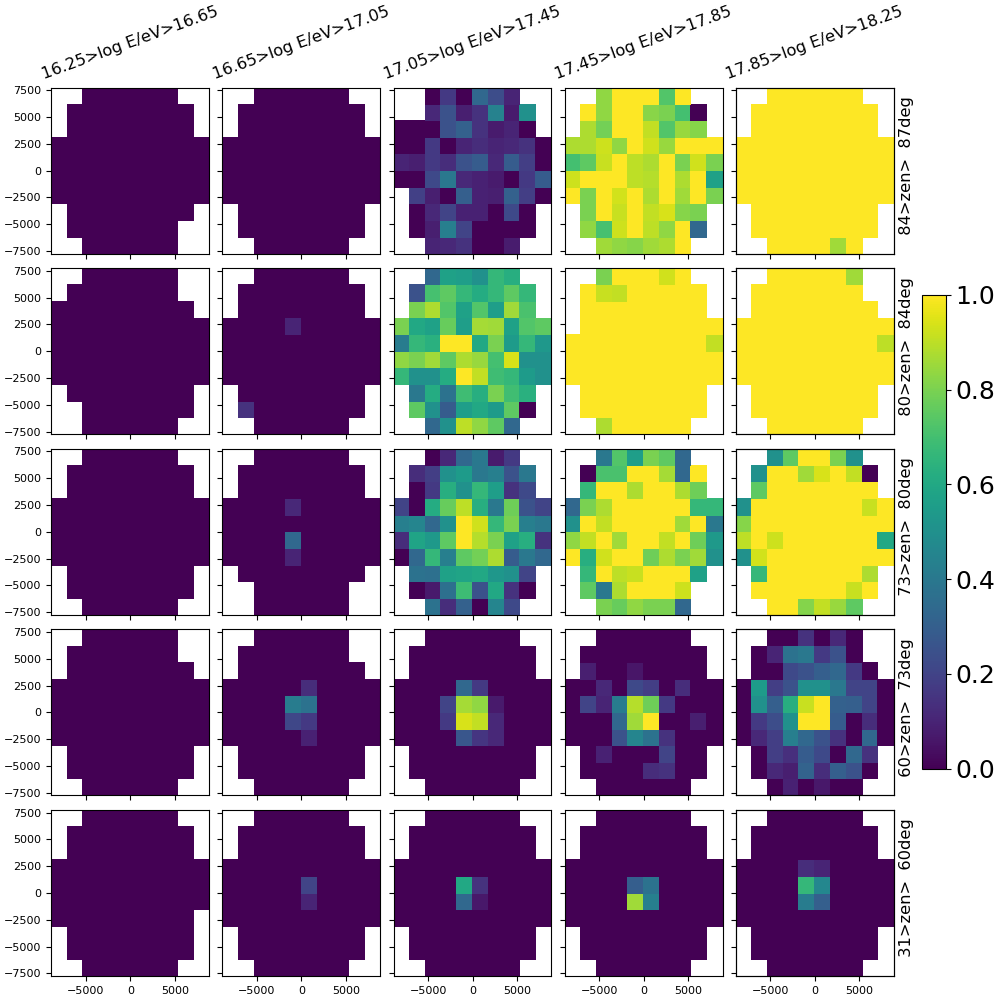}
\caption{Same as Fig.~\ref{fig:heatmap_tri} for the `island' infill layout.}
\label{fig:heatmap_island}
\end{figure}

\begin{figure}[tb]
\includegraphics[width=\linewidth]{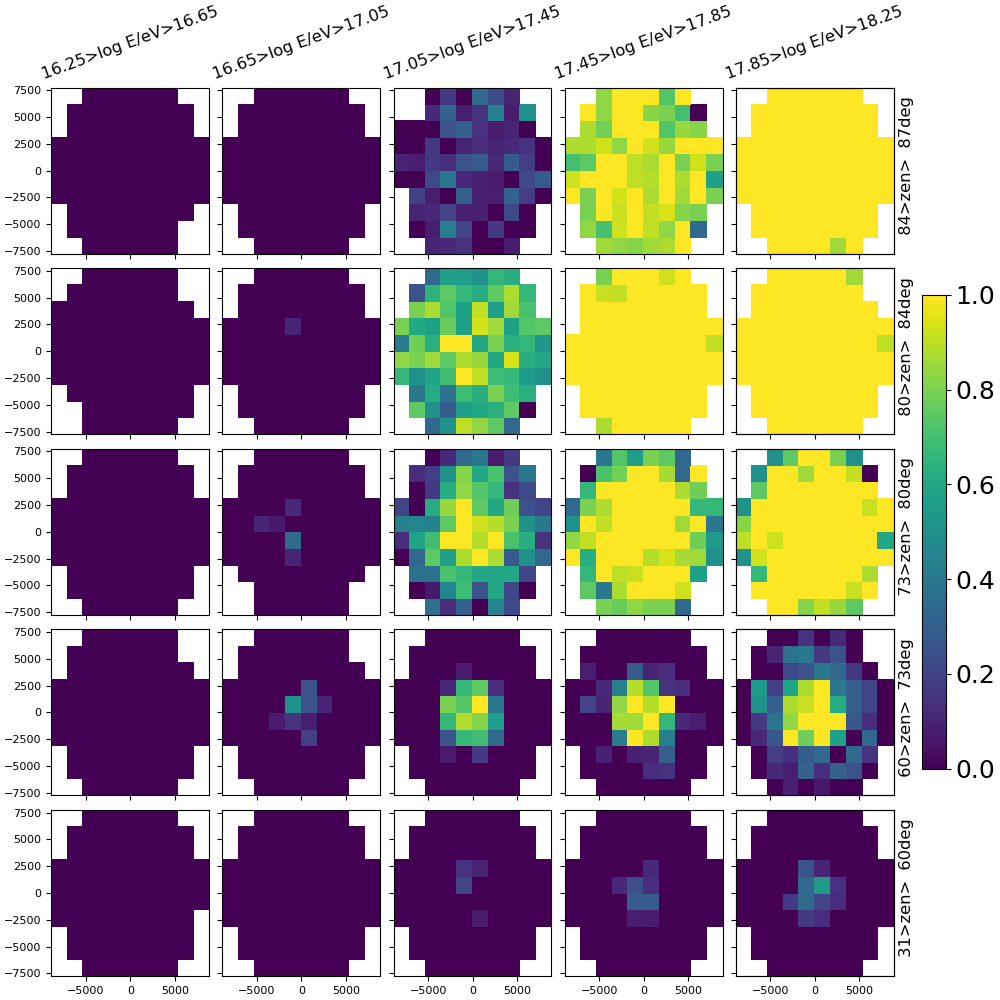}
\caption{Same as Fig.~\ref{fig:heatmap_tri} for the `flower' infill layout.}
\label{fig:heatmap_flower}
\end{figure}

\begin{figure}[tb]
\includegraphics[width=\linewidth]{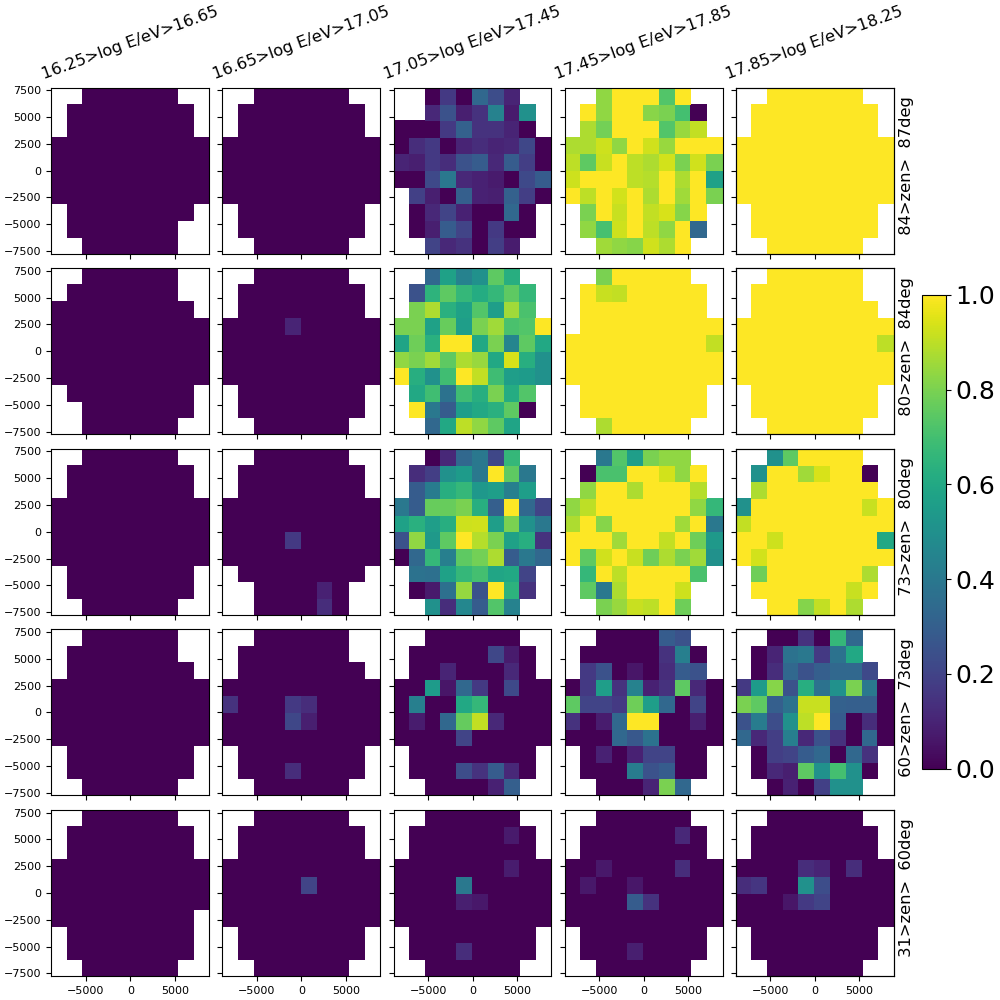}
\caption{Same as Fig.~\ref{fig:heatmap_tri} for the `spiral' infill layout.}
\label{fig:heatmap_spiral}
\end{figure}

Figures~\ref{fig:heatmap_tri}$-$\ref{fig:heatmap_spiral}  complement these results by highlighting the regions where detection occurs in these infill layouts. The figures present heat maps, i.e., the trigger rates computed in pixels of $2.75\,{\rm km}^2$, for 5 energy bins and 5 zenith angle bins, for each infill type. 
The infill regions appear brightly in these maps. The tri infill only triggers 8 panels in the upper right corner. On the other hand, the other patterns present full triggering efficiencies (yellow) on the dense region, over 11 panels on the right-hand side. The dense coverage enables indeed to detect more vertical showers. It appears however that the extension to lower energies remain limited, as was also observed in the left-hand panel of Figure~\ref{fig:infill}. We notice that a for complex pattern such as the spiral one, the effective area calculation for the infill could be intricate and this might not facilitate flux estimates and reconstruction procedures.

\begin{table}
 \centering
 \begin{tabular}{lcrrr}
   \hline \hline
   geometry & step size of coarse & $N_{\rm antenna}$ &$N_{\rm ev}$&  $N_{\rm ev}/N_{\rm antenna}$\\
   &  $\lambda$ [m] &  & [day$^{-1}$] & [day$^{-1}$]  \\
   \hline 
   hex & 1000 &  150 &192.38 & 1.28\\
   tri & 1000 &  241 & 321.77 & 1.33\\
   island & 1000 & 240 & 438.40 & 1.83\\
   flower & 1000 & 264 & 476.43 & 1.80\\
   spiral & 1000 & 235 & 441.15 & 1.88\\
   \hline
   \end{tabular}
   \caption{Characteristics of infill layouts presented in Fig.~\ref{fig:infill_layouts}, all constructed by adding the same number of extra antennas on a coarse `hex' layout with step size $\lambda = 1000\,$m covering a fixed surface of $A_{\rm sim}=204\,$km$^2$. 
   The layouts are obtained by pruning an initial dense tri layout with $N_{\rm ring}=20$ and step size 250\,m. The last column indicates the total number of events detected per day for each layout (see cautionary note on these numbers in Section~\ref{section:geom}).}
  \label{tab:infill}
\end{table}

\section{Conclusion\label{section:conclusion}}

The construction of several very large scale arrays is envisioned and being examined in the ultra-high energy astroparticle community. These projects have increased the demand for efficient computation of air-shower signals over large instrumented areas, in order to explore various layout options. 
In order to answer these demands and circumvent computational costs, we have developed a pruning tool, which consists in running a set of microscopic simulations and interpolate them over a dense, regularly spaced array of detection units, and prune the unnecessary units out of the layout, in order to obtain the shower footprint on a newly shaped layout. 

We showcased our method on the specific case of radio-detection of air-showers. By applying our tool, we assessed the impact of key parameters when choosing a geometrical layout for a specific science case: geometry, size, of single pattern for regular tiling, and different structures and granularity for more complex layouts comprising infill regions. 
We also showed that the tool helps us assess easily the impact of instrumental (triggering) constraints on the event detectability. 

In particular, we discussed the interplay between the energy and inclination of the air-shower on the size of the radio footprint and the intensity of the signal on the ground. We gave estimates of the step size required for homogeneous antennas when targeting a specific region in the primary energy-zenith angle parameter space, under specific triggering conditions. 
These effects were illustrated through the computation of effective areas and event rates over selected layout geometries using the TALE cosmic-ray spectrum \cite{2018ApJ...865...74A}.

Some interesting rule-of-thumb conclusions that can be drawn from the application of the method to radio detection are: i) a hexagonal geometry is more efficient than a triangular geometry, ii) around 1000\,m, the spacing between radio antennas does not drastically change the detection efficiency, iii) for a given number of antennas, adding a granular infill on top of a coarse hexagonal array is more efficient than instrumenting the full array with a less dense spacing. 

We note that a thorough statistical analysis dedicated to estimate the detection rates of an actual array is out of the scope of this paper. The purpose of this article was to demonstrate the usage of the tool, and take advantage from its application to identify layout building strategies for radio detection of ultra-high-energy showers. 

The pruning tool is not restricted to radio arrays at ultra-high-energies. It can in principle be universally applied to explore and optimize arrays of any size, and using any non-imaging detection technique (scintillators, \v{C}erenkov water tanks, radio antennas, etc), changing the measured quantity and the trigger criteria, at the cost of running dedicated microscopic simulations over a dense array for the corresponding detection method. Its limitations is that the spacing examined has to be a multiple of that dense array. Its flexibility to test various layout parameters, instrumental constraints, and physical inputs, and its radical gain in terms of CPU computing time, makes of the pruning method presented here, a promising tool for the community.

\acknowledgments

We thank the GRAND Collaboration for fruitful discussions, and in particular Anne Zilles, Sandra Le Coz and Olivier Martineau for sharing their {\sc HorizonAntenna} response code. This work was supported by the French Agence Nationale de la Recherche ("APACHE" ANR-16-CE31-0001 and PIA ANR-20-IDEES-0002; France), the CNRS Programme IEA Argentine ("ASTRONU", 303475; France), the CNRS Programme Blanc MITI ("GRAND" 2023.1 268448; France), CNRS Programme AMORCE ("GRAND" 258540; France), the Programme National des Hautes Energies of CNRS/INSU with INP and IN2P3 co-funded by CEA and CNES. KK acknowledges support from the Fulbright-France program. Simulations were performed using the computing resources at the CCIN2P3 Computing Centre (Lyon/Villeurbanne – France), partnership between CNRS/IN2P3 and CEA/DSM/Irfu.

\bibliographystyle{elsarticle-num}
\bibliography{layout}

\end{document}